# Highly tunable broadband coherent wavelength conversion with a fiber-based optomechanical system

Xiang Xi[1], Chang-Ling Zou[2,3], Chun-Hua Dong[2,3*], and Xiankai Sun[1*]

[1]Department of Electronic Engineering, The Chinese University of Hong Kong, Shatin, New Territories, Hong Kong

[2]CAS Key Laboratory of Quantum Information, University of Science and Technology of China, Hefei 230026, China

[3]CAS Center for Excellence in Quantum Information and Quantum Physics, University of Science and Technology of China, Hefei, Anhui 230026, China

[*]Corresponding author. Email: chunhua@ustc.edu.cn (C.H.D.); xksun@cuhk.edu.hk (X.S.)

**The modern information networks are built on hybrid systems working at disparate optical wavelengths. Coherent interconnects for converting photons between different wavelengths are highly desired. Although coherent interconnects have conventionally been realized with nonlinear optical effects, those systems require demanding experimental conditions such as phase matching and/or cavity enhancement, which not only bring difficulties in experimental implementation but also set a narrow operating bandwidth (typically in MHz to GHz range as determined by the cavity linewidth). Here, we propose and experimentally demonstrate coherent information transfer between two orthogonally propagating light beams of disparate wavelengths in a fiber-based optomechanical system, which does not require any sort of phase matching or cavity enhancement of the pump beam. The coherent process is demonstrated by phenomena of optomechanically induced transparency and absorption. Our scheme not only significantly simplifies the experimental implementation of coherent wavelength conversion, but also extends the operating bandwidth to that of an optical fiber (tens of THz), which will enable a broad range of coherent-optics-based applications such as optical sensing, spectroscopy, and communication.**

Light at different wavelengths has distinct features in light–matter interactions. Those distinct strengths have been exploited in separate optical systems for various applications, such as ultrasensitive sensing, spectroscopy, metrology, and communication[1-4]. To leverage the strengths of those separate systems and develop optical information networks with increased size and complexity, it is often required to convert photons coherently between disparate wavelengths. The conventional wisdom on wavelength conversion relies on materials' optical nonlinearities[5-7]. Recently, enabled by advanced nanofabrication techniques, a more efficient scheme exploiting photon–phonon interactions in optomechanical structures[8-13] has been intensively explored for achieving coherent wavelength conversion[14-19]. With controllable photon–phonon interactions, the photon conversion efficiency can potentially reach unity by constructing an optomechanical dark state[20]. To date, the strong and tailorable coupling between photons and phonons has greatly facilitated coherent information conversion among optics, mechanics, and other physical degrees of freedom in both the classical and quantum regimes[21-28]. Despite great success, the existing wavelength conversion systems by using either nonlinear optical effects or photon–phonon interactions usually require strict phase matching between traveling photons and phonons[17-19] or strong cavity enhancement of resonant photons at both wavelengths[14-16,18,20]. These stringent requirements add to difficulties and limitations in experimental implementation, such as limited selection of operating wavelengths and narrow operating bandwidth. As a result, the traditional schemes for wavelength conversion are impractical because of the incompatibility with other wavelength-independent optical components. Therefore, strategies without the above difficulties and limitations for universal coherent conversion of photons between disparate optical systems are being sought.

Here, we experimentally demonstrate coherent optical wavelength conversion between two orthogonally propagating light beams of disparate wavelengths in a fiber-based optomechanical system as shown in Fig. 1(a). In the system, a high-quality mechanical mode serves as a coherent link between the pump and probe photons at different wavelengths. The information carried by the pump light traveling in the fiber core is coherently transferred to the probe light of a cavity mode traveling orthogonally in the fiber cladding. Since coherent signal conversion via the photon–phonon interactions can be identified by the phenomena of optomechanically induced transparency or electro-optomechanically induced transparency[15,25,26], here we prove coherent signal conversion between different wavelengths by demonstrating the symbolic phenomenon of optomechanically induced transparency and absorption, which is enabled by the broadband nonresonant pump photons. Compared with previous methods, our scheme features the following advantages: Owing to the high quality factors of the mechanical mode (~16,000) and the optical probe mode (~1.3 × $10^8$), the pump light that actuates the mechanical motion does not need cavity enhancement or phase matching, thus avoiding the previously mentioned difficulties and limitations. The operating bandwidth of the pump light is now the working wavelength range of the optical fiber, which is typically tens of terahertz and a 4-order-of-magnitude enhancement from that of the existing schemes. Additionally, in our experimental configuration the pump and probe optical beams are spatially isolated, which can avoid the undesired nonlinear effects that usually exist in the other systems. This can largely reduce signal crosstalk and noise introduced by the strong pump beam.

Figure 1(a) is a conceptual illustration of the fiber-based optomechanical system. It is made of a standard single-mode fiber with part of its cladding fused to form a microbottle cavity. The microbottle cavity can support several groups of mechanical modes [Fig. 1(b)], where the radial-contour modes (RCMs, denoted as $R_{k,0}$) were employed in our experiments. Meanwhile, the microbottle cavity can also support abundant high-quality optical whispering-gallery cavity modes traveling in the equator of the microbottle, which are coupled in and out via a tapered fiber placed in proximity of the microbottle along the transverse direction [Fig. 1(a)]. These whispering-gallery cavity modes have strong optomechanical coupling with the RCMs, and many of them can be selected as the probe mode $a$.

When an optical pump field with mixed components of $a_1$ (control) and $a_{in}$ (signal) is launched into the core of the fiber and propagates along the $z$ direction, it will exert an oscillating optical force in the fiber, causing a mechanical vibration of the microbottle cavity [Figs. 1(a) and 1(b)]. Such mechanical vibrations can be resonantly amplified if the frequency difference $\Omega_m$ between $a_1$ and $a_{in}$ matches the frequency of one of the mechanical modes of the microbottle cavity. The actuated mechanical vibrations in turn modulate the intracavity modal field $a$ [Figs. 1(a) and 1(b)] through the optomechanical coupling, producing a converted sideband signal $a_{cs}$ with a new frequency component in the output of the probe control light $a_2$. With such a device system, the information carried by the input signal $a_{in}$ is converted to the output field $a_{cs}$ with mechanical vibration (or phonon) as a coherent link [Figs. 1(b) and 1(c)]. Compared with conventional optomechanical methods which have stringent requirements, the only requirement here is that the pump light should be coupled into the fundamental mode of the optical fiber to ensure a broad operating wavelength range.

When the optical pump beam consists of two components $a_1 e^{-i\omega_1 t}$ and $a_{\mathrm{in}} e^{-i(\omega_1+\Omega)t}$, the global amplitude $x(t)$ of mechanical modes is determined by an overlap integral of the spatial distributions of the mechanical mode $\mathbf{u}_m(\mathbf{r})$ and optical force density $\mathbf{F}(\mathbf{r})$, $G_m = \int \mathbf{u}_m(\mathbf{r})\cdot\mathbf{F}(\mathbf{r})\mathrm{d}V$ (see Supplementary Sec. 1.3)[29,30]. In a frame rotating at $\omega_1$, the global effective optical force exerted by the pump fields on the mechanical mode is expressed as $F_{\mathrm{eff}}(t) = G_m a_1 a_{\mathrm{in}} e^{-i\Omega t}$ (see Supplementary Sec. 1). Therefore, the dynamics of mechanical amplitude actuated by the pump fields in the fiber core is governed by

$$m_{\mathrm{eff}}\left(\ddot{x}(t) + \Gamma_m \dot{x}(t) + \Omega_m^2 x(t)\right) = F_{\mathrm{eff}}(t),\tag{1}$$

where $m_{\mathrm{eff}}$ is the effective mechanical mass, $\Omega_m$ and $\Gamma_m$ are the eigenfrequency and damping rate of the mechanical mode, respectively (see Supplementary Sec. 2). Meanwhile, the dynamics of the optical probe field $a$ mediated by the mechanical motion $x(t)$ is governed by

$$\dot{a}(t) = \left(i\Delta - \kappa/2\right)a(t) + iGx(t)a(t) + \sqrt{\kappa_{\mathrm{ex}}}\,a_2,\tag{2}$$

where the optomechanical coupling rate $G$ is defined as the optical resonant frequency shift per unit mechanical displacement $G = \partial\omega/\partial x$. In the above equation, $\Delta = \omega_2 - \omega_{\mathrm{cav}}$ is the frequency detuning of the control field $a_2$ to the cavity resonance. The total cavity decay rate is $\kappa = \kappa_{\mathrm{ex}} + \kappa_{\mathrm{in}}$, where $\kappa_{\mathrm{ex}}$ is the taper–cavity coupling rate and $\kappa_{\mathrm{in}}$ is the intrinsic cavity damping rate. Different from the conventional cavity optomechanical scheme[14,20], here the optical pump beams ($a_1$ and $a_{\mathrm{in}}$) are single-pass traveling laser fields, so the mechanically induced phase change of the optical pump does not exert backaction on the actuated mechanical modes. Besides, the effect of optical force on mechanical motion by the control field $a_2$ is orders of magnitude weaker than that by the pump field, so the dynamic backaction by the control field $a_2$ is also negligible. Therefore, Eqs. (1) and (2) are sufficiently accurate to describe the wavelength conversion dynamics in Figs. 1(b) and 1(c). By solving these equations in the frequency domain (see Supplementary Sec. 2), the frequency-dependent converted signal field $a_{\mathrm{cs}}$ that is optomechanically converted from the signal field $a_{\mathrm{in}}$ is expressed as

$$a_{\mathrm{cs}} = \frac{\sqrt{\kappa_{\mathrm{ex}}}\,G G_m a_1 \bar{a}}{-2m_{\mathrm{eff}}\Omega_m \left[-i(\Omega-\Omega_m) + \Gamma_m/2\right]\left[-i(\Delta+\Omega) + \kappa/2\right]}\, a_{\mathrm{in}}.\tag{3}$$

Here $\overline{a} = \sqrt{\kappa_{\text{ex}}} a_2 / (\kappa/2 - i\Delta)$ is the steady state of intracavity field. Therefore, the total output signal would be

$$a_{\text{total}}(t) = a_2 e^{-i\omega_2 t} - (\sqrt{\kappa_{\text{ex}}} \overline{a} e^{-i\omega_2 t} + a_{\text{cs}} e^{-i(\omega_2 + \Omega)t}). \qquad (4)$$

When $\Omega$ matches the eigenfrequency of the mechanical mode $\Omega_m$, it is possible to obtain efficient wavelength conversion from the input signal $a_{\text{in}}$ to the output signal $a_{\text{cs}}$. According to Eq.(3), low mechanical and optical damping rates $\Gamma_m$ and $\kappa$ can enhance the conversion efficiency.

Figure 2 shows the fabricated device, where the cladding of the fiber is slightly fused to reduce the diameter at the two neck positions and form a bottle-like microstructure [Fig. 2(b)]. Such a bottle-like microstructure forms optical and mechanical energy potentials that support high-quality optical probe modes and mechanical modes [Fig. 1(b) and 2(c)]. At first, we characterized the mechanical modes that can be utilized for coherent wavelength conversion with an experimental setup in the ambient environment as shown in Fig. 3(a). The optical performance of the cavity modes of the microbottle cavity was characterized by measuring the transmission spectrum of the probe laser. Figure 3(b) shows an optical resonance with a loaded quality factor as high as $1.3 \times 10^8$ at 193.03 THz, indicating that the optical damping rate $\kappa$ is as small as 1.5 MHz. Next, we sent the pump light containing a pump control field $a_1$ (frequency $\omega_1 = 193.4$ THz, power 156 mW) and an intensity-modulation-produced signal field $a_{\text{in}}$ into the fiber core to actuate the mechanical motion of the microbottle cavity. Meanwhile, we sent a control laser field $a_2$ (power 40 μW) with $\Delta \approx -\kappa/2$ into the tapered fiber that was coupled to the microbottle cavity. As discussed above, the the optomechanical transduction process would generate the converted field $a_{\text{cs}}$ of new frequency. Therefore, the mechanical modes can serve as a coherent link for converting the input field $a_{\text{in}}$ to the field $a_{\text{cs}}$. A high-speed photodetector was used to record output signal $a_{\text{total}}$ representing the beating signal between the converted output signal $a_{\text{cs}}$ and the transmitted probe light $a_2$. Figure 3(c) plots the measured $|S_{21}|$ spectrum of the beating signal $|a_{\text{total}}|^2$, which shows the frequency dependence of the wavelength conversion process. It is clear that there exist several peaks at different frequencies which are orders of magnitude higher than the noise floor, indicating efficient generation of the optical field $a_{\text{cs}}$ at those frequencies. With the aid of finite-element simulation, it was identified that the appearance of the first four peak windows ①−④ for wavelength conversion, whose quality factors $Q_m$ range from 1,500 to 5,000, are attributed to the first-order mechanical RCMs $R_{1,0}$ with different axial wavenumbers [Fig. 3(e)]. The peak ⑤ is attributed to the second-order mechanical RCM $R_{2,0}$ [Fig. 3(e)]. Its quality factor is as high as 16,000 in air, leading to a $Q_m \cdot f_m$ product of $1.43 \times 10^{12}$ which is among the best values obtained from similar structures[31]. The high quality factors of the mechanical and optical modes of the microbottle cavity can compensate for the lacking of cavity enhancement for the pump light, making the wavelength conversion experimentally feasible. We measured the wavelength conversion process for different devices, and found that the conversion efficiency from $a_{in}$ to $a_{cs}$ is almost independent of the wavelength of the pump light over a broad wavelength range, as shown in Fig. 3(d). The wavelength of the pump beam adopted in this experiment was limited by the working bandwidth of the erbium-doped fiber amplifier. The device can actually work in a much larger bandwidth that covers the entire wavelength range from the O band to L band. With a modified diameter of the fiber core, the working bandwidth can be extended further to the visible or mid-infrared optical bands. Note that the $|S_{21}|$ spectrum was measured under the condition of $\Delta \approx -\kappa/2$, and thus the conversion efficiency cannot be accurately determined as small sidebands of other frequency orders also exist. Accurate calibration of the conversion efficiency requires operation in the resolved-sideband regime with $|\Delta| \gg \kappa$, as discussed below.

The coherent wavelength conversion via photon–phonon interaction in cavity optomechanics is symbolized by the phenomenon of optomechanically induced transparency (OMIT) in the optical domain[15,18,22], and electro-optomechanically induced transparency in an extended frequency range covering both the microwave and optical domains[25,26]. According to Eq. (3), the maximal wavelength conversion efficiency can be achieved under the condition of $\Delta = -\Omega_m$. To demonstrate the coherent wavelength conversion process with maximal conversion efficiency, we also conducted experiments that exhibit the key features of OMIT. The OMIT in our scheme can be obtained by using nonresonant pump light without the bandwidth limitation, so it is termed broadband optomechanically induced transparency (BOMIT). We sent the control laser $a_2$ (frequency $\omega_2$, and $\Delta = -\Omega_m$) and its phase-modulated sideband $a_p$ (frequency $\omega_p = \omega_2 + \Omega$) into the tapered fiber [Fig. 4(a)]. Now, in the system there are two pathways for generating intracavity photons: one is through the optomechanical conversion by the nonresonant pump light propagating in the fiber core, and the other is the directly injected probe field $a_p$. When these two optical fields interfere destructively (constructively), a narrow transparency (absorption) window appears in the cavity transmission spectrum. This process is governed by the coupled equations of motion for the complex intracavity optical field $A^-$ and mechanical amplitude $x$, with $A^-$ expressed as

$$A^- = \frac{GG_m a_1 \overline{a} a_{\text{in}} e^{i\phi}}{-2m_{\text{eff}}\Omega_m \left[-i(\Omega - \Omega_m) + \Gamma_m/2\right]\left[-i(\Delta + \Omega) + \kappa/2\right]} + \frac{\sqrt{\kappa_{\text{ex}}} a_{\text{p}}}{-i(\Delta + \Omega) + \kappa/2}, \tag{5}$$

where $\phi$ is the modulation phase shift of the pump beam that can be set and controlled by a phase shifter in the experiment (see Supplementary Sec. 3). In Eq. (5), the first term is contributed by the optomechanical conversion from the input signal $a_{\text{in}}$, and the second term is contributed by the direct injection of the probe field $a_{\text{p}}$.

The phenomenon of BOMIT was demonstrated with the aid of the high-quality-factor mechanical mode $R_{2,0}$. We measured the power transmission spectra of the microbottle cavity by using a heterodyne detection setup as shown in Fig. 4(b). We scanned the modulation frequency $\Omega$ and observed a narrow transparency dip in the sideband transmission spectrum of the probe light [Fig. 4(c), upper panel]. The transparency dip is located at frequency $\Omega = \Omega_m$, which confirms its origin of destructive interference between the optomechanically converted photons and the directly injected probe photons. The transparency dip turned into a peak as the modulation phase $\phi$ for the pump laser was varied by $\pi$ [Fig. 4(c), lower panel]. This is because the interference now becomes constructive, leading to the phenomenon of broadband optomechanically induced absorption (BOMIA). When the modulation phase $\phi$ is between 0 and $\pi$, the transmission spectra exhibit a Fano-resonance-like line shape as shown in Fig. 4(d).

The nearly total destructive interference represented by the transparency dip in Figs. 4(c) and 4(d) indicates that the optomechanically converted optical field $a_{\text{cs}}$ has the same amplitude as the probe field $a_{\text{p}}$. In this case, the external convertion efficiency $(a_{\text{cs}}/a_{\text{in}})^2$ is found to be ~$3 \times 10^{-5}$. This efficiency is lower than some of the existing works[14,20,32], because we used a nonresonant propagating pump to actuate the mechanical motions, while the conventional schemes employ an optical cavity to enhance the optomechanical interaction between the pump light and mechanical modes. Despite a lower conversion efficiency, our nonresonant scheme extends the operating bandwidth of the pump light to tens of terahertz, which is a 4-order-of-magnitude enhancement from that of the existing schemes. Another important reason for the low efficiency in the current system is the small $G_m$, because of the small overlap between the mechanical mode $\mathbf{u}_m(\mathbf{r})$ and the optical force density $\mathbf{F}(\mathbf{r})$ (See Supplementary Fig. S1). The $G_m$ can be enhanced by redesigning the structure of the optical fiber, e.g., by increasing the size of the fiber core and reducing the size of the fiber cladding.

The amplitude of the optomechanically scattered sideband can also be tuned independently by the intensity of the pump light, achieving a tunable transparency (or absorption) window. Figure 5(a) shows the power transmission spectra of the probe light under different power levels of the pump light measured in the case of BOMIT. Since the optomechanically converted optical field interferes destructively with the probe field, the transparency dip first decreases and then increases as the pump power increases, with a quadratic tendency as shown in Fig. 5(b). Figure 5(c) shows the power transmission spectra of the probe light under different power levels of the pump light measured in the case of BOMIA. Since the optical interference is now constructive, the absorption peak keeps increasing as the pump power increases, also with a quadratic tendency as shown in Fig. 5(d). These experiments agree well with the prediction of Eq. (5), and indicate that the coherent wavelength conversion efficiency can be tuned by the power of pump light. It was found that, as the pump power increases, the transparency dip and absorption peak shift slightly to the higher frequency side. This is attributed to the stress induced by the nonoscillating optical force of the pump light[33].

In conclusion, we have proposed a fiber-based optomechanical system for coherent wavelength conversion and experimentally demonstrated the viability through the phenomena of broadband optomechanically induced transparency and absorption with high tunability. This scheme largely relieves the experimental requirements for the pump light, and enables operation over a broad wavelength range the same as that of a single-mode fiber, which is typically tens of terahertz and a 4-order-of-magnitude enhancement from that the existing schemes. The high tunability and broad operating wavelength range feature clear advantages for applications of wavelength conversion between photons in disparate optical systems, and provides new opportunities for building optical information networks with extended sizes and complexity. Besides, considering that the mechanical modes of optical fibers are employed for distributed sensing[34,35], the fiber-based optomechanical system in this work can also be exploited for enhanced distributed fiber sensing, because the combination of its high-quality mechanical and optical cavity modes can provide unprecedently higher sensitivity and smaller footprint than those of the traditional fiber sensing schemes. With these advantages, our device system will play an important role in modern optical signal processing and communications.

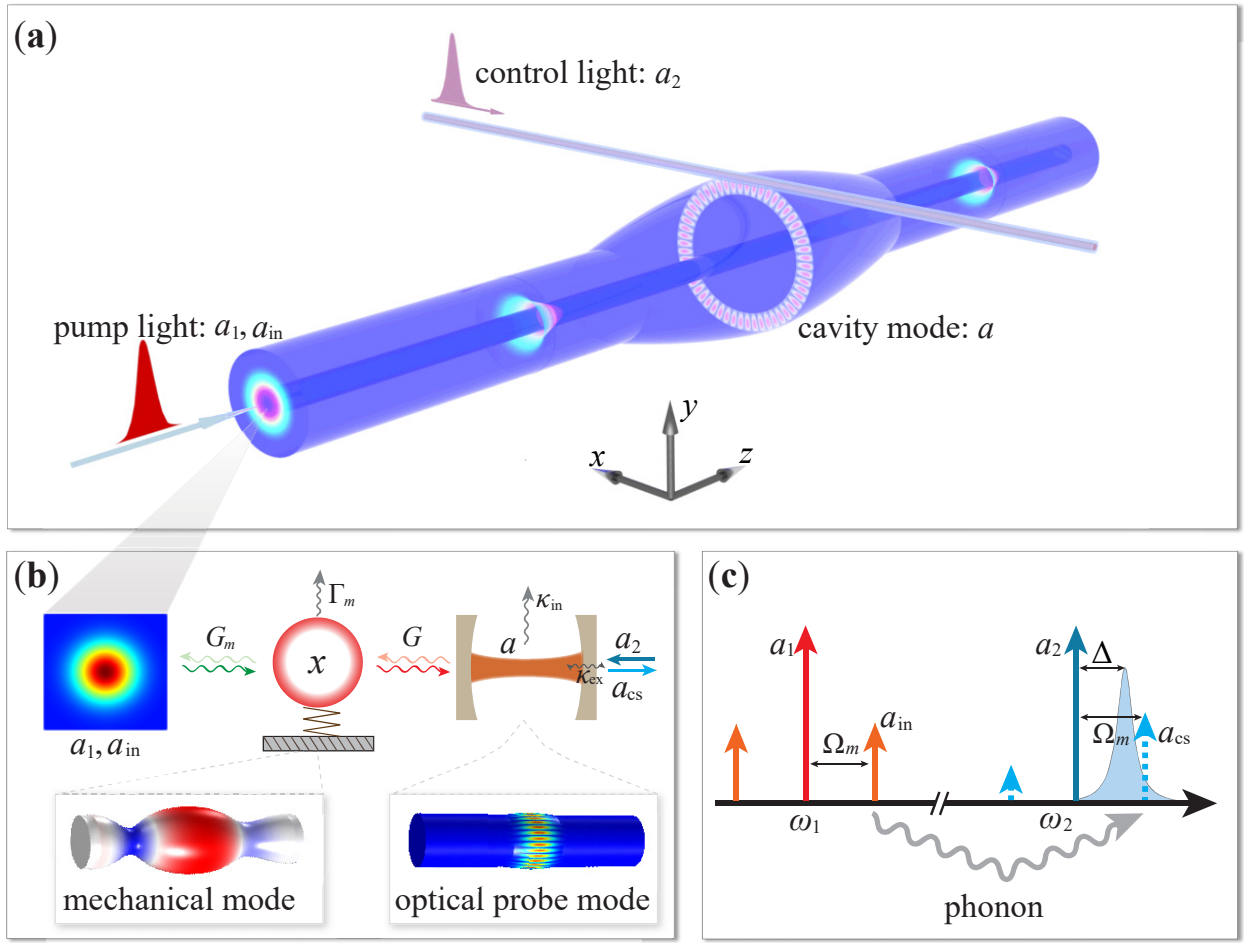

**Fig. 1. Concept of coherent wavelength conversion with a fiber-based optomechanical system.**
(**a**) Schematic of the coherent wavelength conversion process. A pump beam propagating in fiber core exerts an optical force in the transverse (*x-y*) direction to excite the mechanical motion. The mechanical motion modulates a probe field traveling in the fiber cladding in the transverse plane to generate an optical field of new wavelength. (**b**) Diagram of the coherent wavelength conversion from a longitudinally propagating Gaussian mode to a transversely resonating cavity mode. The single-pass pump field (control: $a_1$, signal: $a_{in}$) and probe cavity mode ($a$) are coupled to the same mechanical mode ($x$) with coupling strengths $G_m$ and $G$, respectively. A control mode $a_2$ is coupled into the cavity with coupling rate $\kappa_{ex}$. The intrinsic damping rates of the optical probe mode and the mechanical mode are $\kappa_{in}$ and $\Gamma_m$, respectively. Shown at the bottom left is the simulated displacement field of a mechanical radial-contour mode. Shown at the bottom right is the simulated optical probe field in a whispering-gallery mode traveling in the fiber cladding. (**c**) Schematic showing the relevant optical frequencies involved in the wavelength conversion process. The information carried in $a_{in}$ is converted through optomechanical interaction to a new frequency $a_{cs}$.

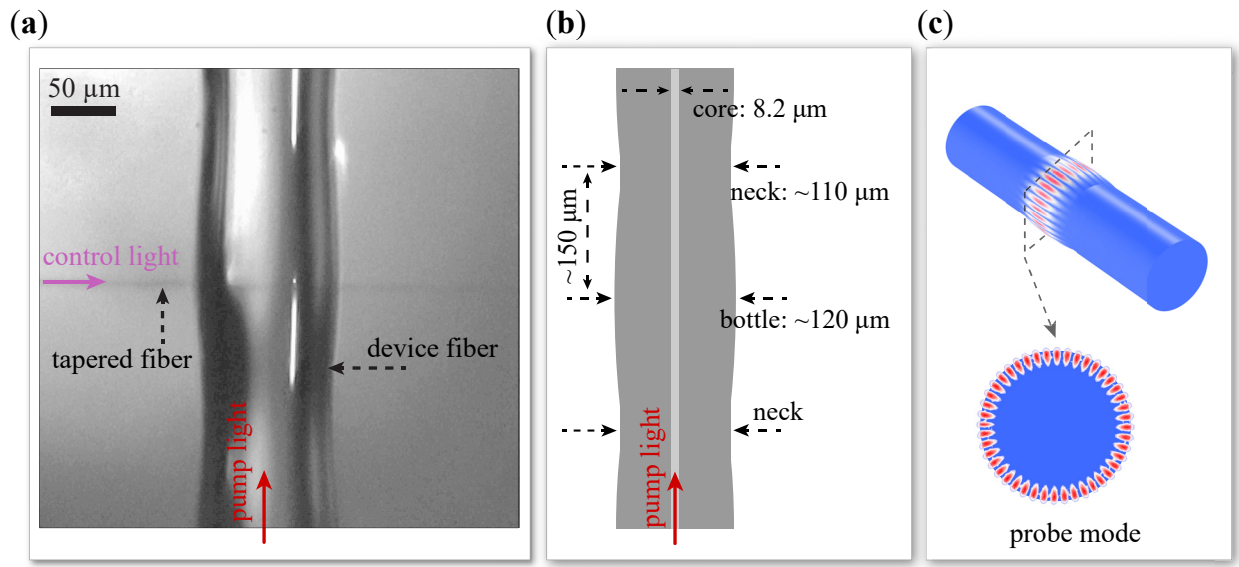

**Fig. 2. Configuration and geometry of the device used in experiment.** (**a**) Optical microscope image of the microbottle part of the fabricated device fiber and the tapered fiber. The pump light propagating in the core of the device fiber is used to actuate the mechanical modes of the microbottle cavity. (**b**) Geometry of the fabricated device fiber in (**a**). At the two neck positions, the size of the fiber cladding is reduced by laser fusing to create a microbottle-like cavity in the middle. (**c**) Illustration of the optical probe mode supported by the microbottle cavity, which can be modulated by the actuated mechanical motion. The probe mode is accessed by the probe light propagating in the tapered fiber in (**a**) that is evanescently coupled with the microbottle cavity.

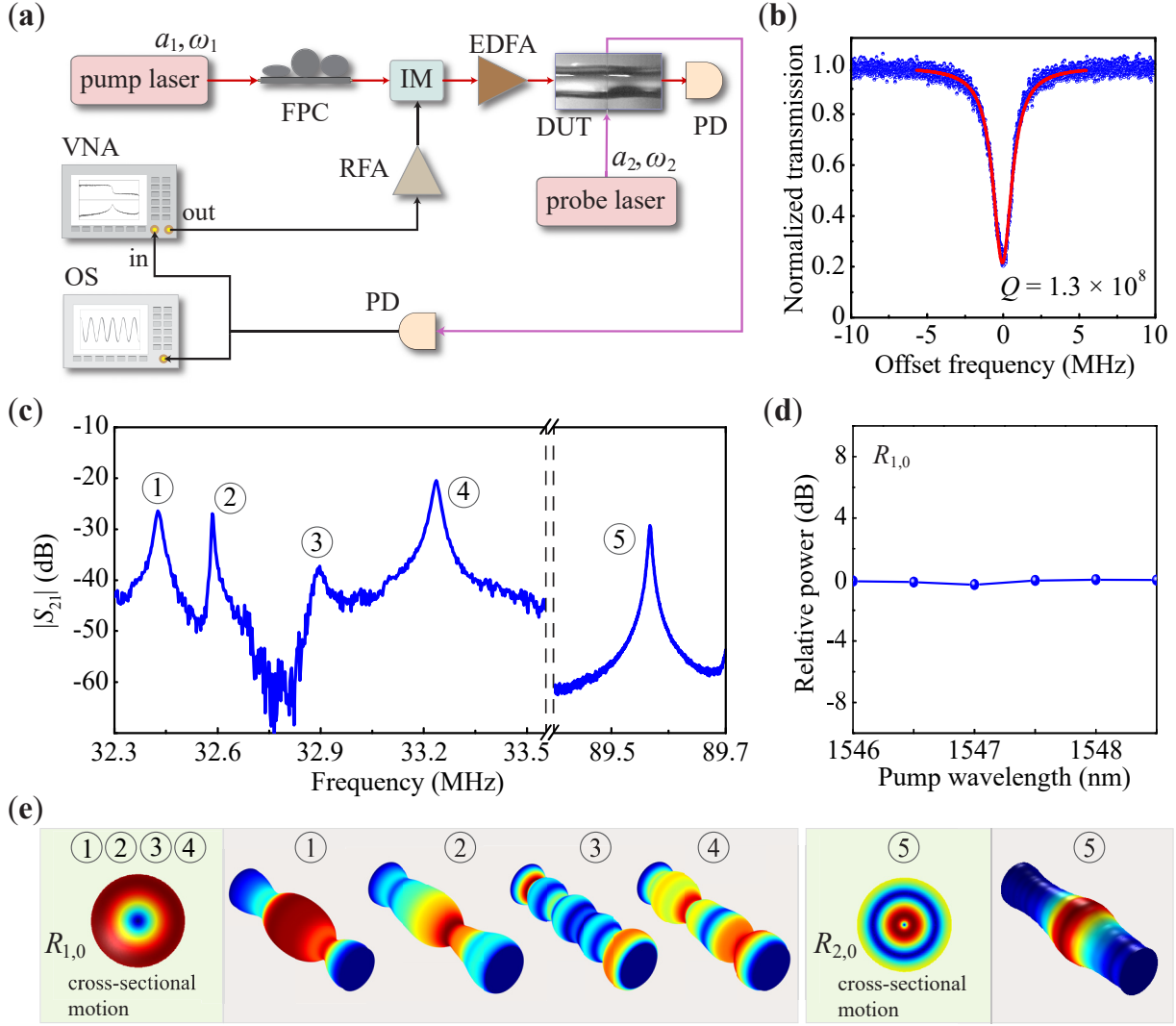

**Fig. 3. Experimental demonstration of broadband coherent wavelength conversion.** (**a**) Experimental setup. A pump laser field $a_1$ (frequency 193.4 THz) was intensity-modulated to generate an input signal $a_{in}$. A probe control laser field $a_2$ was sent via a tapered fiber into the microbottle cavity, which was modulated by the mechanical motion of the fiber to produce a new converted signal $a_{cs}$. DUT, device under test; IM, intensity modulator; EDFA, erbium-doped fiber amplifier; FPC, fiber polarization controller; OS, oscilloscope; PD, photodetector; RFA, radio-frequency amplifier; VNA, vector network analyzer. (**b**) Optical transmission spectrum of the probe light showing a resonance with an optical quality factor of $1.3 \times 10^8$ at 193.03 THz. (**c**) Measured $S_{21}$ power spectrum showing the mechanical driven response. This spectrum provides the relative strength of the converted signal $a_{cs}$ through optomechanical interactions. (**d**) Measured relative optical power converted from pump light of different wavelengths via one of the mechanical modes $R_{1,0}$, which indicates that the coherent wavelength conversion process is independent of the wavelength of the pump light. (**e**) Simulated mechanical modal profiles for the five peaks shown in the measured $S_{21}$ spectrum in (**c**).

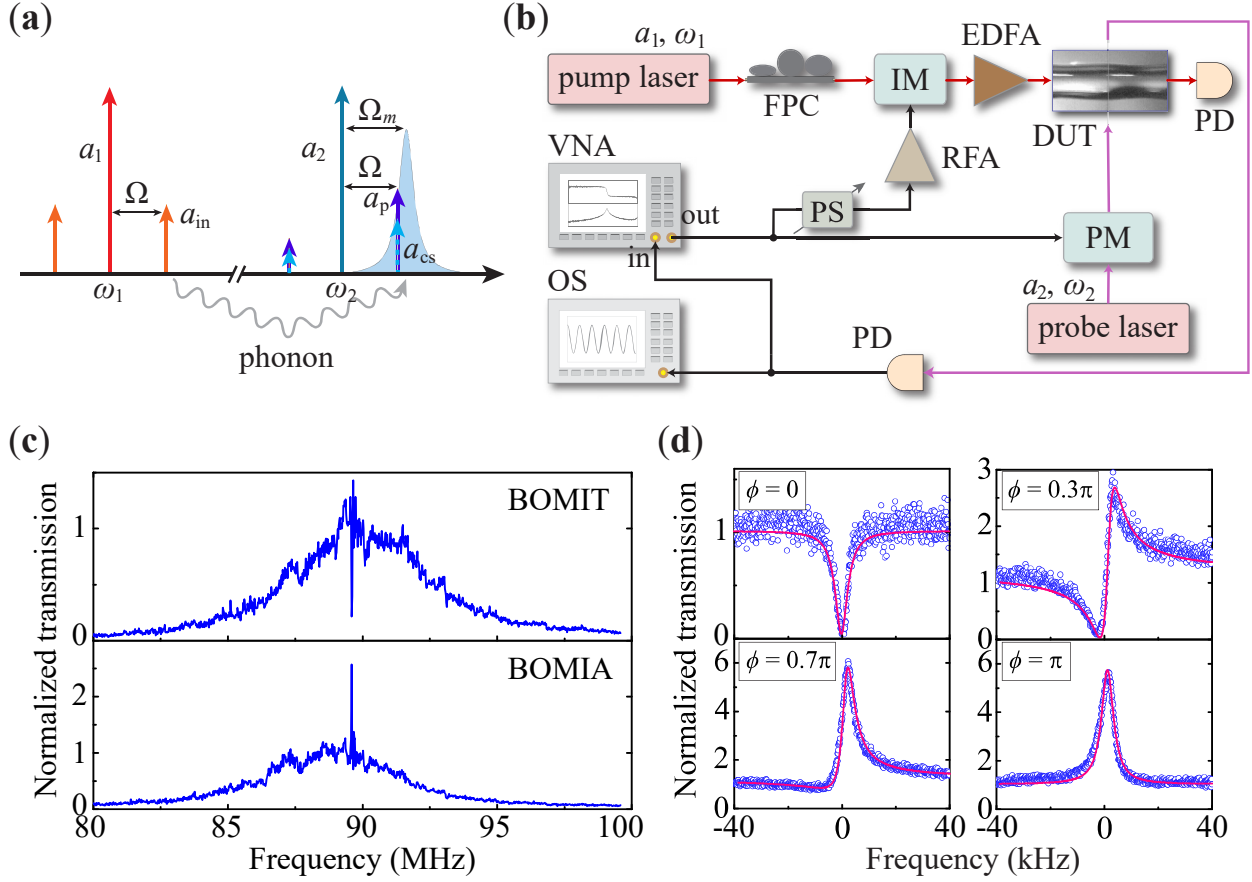

**Fig. 4. Experimental demonstration of broadband optomechanically induced transparency and absorption.** (**a**) Schematic showing the relevant optical frequencies involved in the BOMIT process. In addition to the pump fields ($a_1$ and $a_{in}$) for phonon actuation, a probe control field ($a_2$) at frequency $\omega_2$ was red-detuned from the cavity mode at a frequency $\Omega_m$. Another probe field $a_p$ (violet solid arrow) was sent to obtain the intracavity optical power spectrum. (**b**) Experimental setup for measuring the BOMIT. The probe light was generated by phase-modulating the control field $a_2$ at a frequency $\Omega$. The radio-frequency (RF) source for modulating the pump control $a_1$ and probe control $a_2$ is from the same VNA. PM, phase modulator; PS, phase shifter. (**c**) Transmission spectra of the probe light showing the BOMIT and BOMIA. Whether it exhibits BOMIT or BOMIA is determined by the RF phase $\phi$ between the driving field $a_{in}$ and the probe field $a_p$. (**d**) Transmission spectra of the probe light for $\phi = 0$, $0.3\pi$, $0.7\pi$, and $\pi$. When $\phi \neq 0$ or $\pi$, the resonance exhibits a Fano line shape.

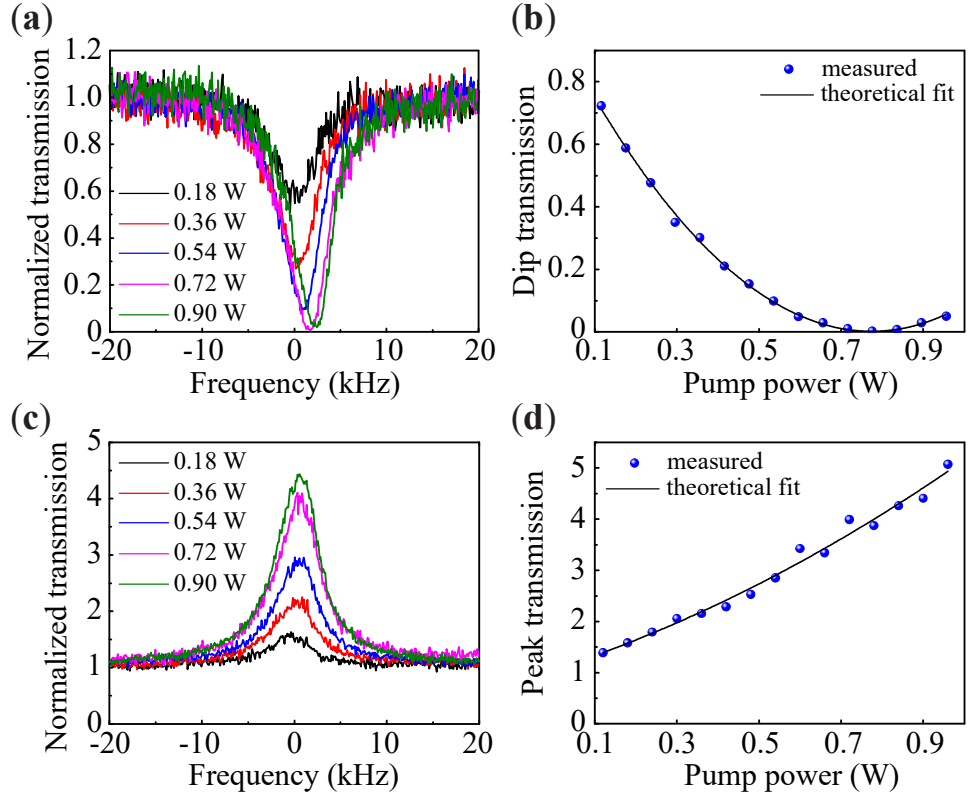

**Fig. 5. Experimental demonstration of tunability of optomechanically induced transparency and absorption by pump power.** (**a**) Normalized transmission spectra of optomechanically induced transparency measured under various pump power levels. (**b**) Transmission at the dip of the spectra, i.e., the OMIT signal strength, in (**a**) as a function of the pump power. The black line is a theoretical fit of the experimental data. (**c**) Normalized transmission spectra of optomechanically induced absorption measured under various pump power levels. (**d**) Transmission at the peak of the spectra, i.e., the OMIA signal strength, in (**c**) as a function of the pump power. The black line is a theoretical fit of the experimental data.